\documentclass[namedreferences]{solarphysics}
\usepackage[hyperref,optionalrh,solaromanenum]{spr-sola-addons} % For Solar Physics 
\usepackage{graphicx}                    % For eps figures, newer & more powerfull
\usepackage{color}                       % For color text: \color command
\newcommand{\etal}{{\it et al.}}

\begin{document}

\begin{article}

\begin{opening}

\title{Ripples and Rush-to-the-Poles in the photospheric magnetic field}
% variants

%%%%%%%%%%%%%%%%%%%%%%%%%%%%%%%%%%%%%%%%%%%%%%%%%%%
%% Authors Names
%
% \author[addressref={},corref,email={}]{\inits{}\fnm{}\lnm{}\orcid{}}
\author[addressref={aff1},corref,email={elenavernova96@gmail.com}]{\inits{E.S.}\fnm{E.S.}~\lnm{Vernova}\orcid{0000-0001-8075-1522}}
\author[addressref=aff1]{\inits{M.I.}\fnm{M.I.}~\lnm{Tyasto}}\sep
\author[addressref={aff2},email={d.baranov@bk.ru}]{\inits{D.G.}\fnm{D.G.}~\lnm{Baranov}\orcid{0000-0003-2838-8513}}\sep

%%%%%%%%%%%%%%%%%%%%%%%%%%%%%%%%%%%%%%%%%%%%%%%%%%%
%% Runningheads
%
\runningauthor{Vernova et al.}
\runningtitle{Ripples and Rush-to-the-Poles}

%%%%%%%%%%%%%%%%%%%%%%%%%%%%%%%%%%%%%%%%%%%%%%%%%%%
%% Affilations 
%% id shold be the same with \author addressref value.

\address[id={aff1}]{IZMIRAN, SPb. Filial, Laboratory of Magnetospheric
Disturbances, St. Petersburg, Russia}
\address[id={aff2}]{Ioffe Institute, St. Petersburg, Russia}
%%%%%%%%%%%%%%%%%%%%%%%%%%%%%%%%%%%%%%%%%%%%%%%%%%%
%%% Abstract 
%\begin{abstract}
%\end{abstract}
\begin{abstract}
The distribution of magnetic fields of positive and negative polarities over the surface 
of the Sun was studied on the basis of synoptic maps presented by the NSO/Kitt Peak (1978--2016). 
To emphasize the contribution of weak fields the following transformation of synoptic maps 
was made: for each synoptic map only magnetic fields with modulus less than 5\,G ($|B|\leq 5$\,G) 
were left unchanged while  larger or smaller fields were replaced by 
the corresponding threshold values +5\,G or $-5$\,G. Cyclic variations of the magnetic field 
polarity have been observed associated with two types of magnetic field flows in the photosphere. 
Rush-to-the-Poles (RTTP) appear near the maximum of solar activity and have the same sign as the following 
sunspots. The lifetime of RTTP is $\sim 3$~yrs, during which time they drift from latitudes 
$30^\circ$--$40^\circ$ to the pole, causing the  polarity change of the Sun’s polar field. 
Our aim is the study of another type of variations which has  the form of series of flows with  individual 
flows of 0.5--1~yr  with alternating polarity. These flows called “ripples” by \citet{ulri} are located in time 
between two RTTP and drift from the equator to the latitudes of $\sim 50^\circ$. 
The period of variation of ripples was shown to be 1.1~yr for the northern hemisphere 
and 1.3~yr for the southern hemisphere. It was found that the amplitude of variation was higher for the time intervals 
where the polar field had a positive sign. Within the same flow, fields of positive and negative 
signs developed in anti-phase. Two types of flows -- RTTP and ripples -- together formed a unique structure
which had close connection to the magnetic solar cycle.

\end{abstract}

%%%%%%%%%%%%%%%%%%%%%%%%%%%%%%%%%%%%%%%%%%%%%%%%%%%
%% Keywords
%
%\keywords{}

\end{opening}
%-------------------------------------------------

\sloppy

\section{Introduction}

Magnetic field of the Sun governs all manifestations of the solar activity (SA). Magnetic field groups of different magnitudes from the most strong magnetic fields to the background magnetic fields are connected with certain solar phenomena.
Cyclic changes of the solar activity reflect the periodic change of Sun’s magnetic field. The magnetic fields follow the 22-year magnetic cycle: the law of the change of the polarity which manifests itself in the change of the sign of the polar field near the maximum of SA and in the change of the sign of the leading and the following sunspots near the minimum of SA. Important feature of the magnetic field cycles is the restructuring  of the field  distribution over the Sun’s surface.
The variation of SA with the 11-year cycle (Schwabe law) manifests itself in the solar activity distribution as the Maunder butterflies. 

Numerous studies deal with the features of magnetic fields distribution over the Sun’s surface, in particular with the asymmetry of the distribution. Such phenomena as  active longitudes (\citealp{gaiz}; \citealp{bai}; \citealp{biga} and the references therein), and north-south asymmetry (\citealp{bal}; \citealp{deng} and references therein)  play an important role in the development of solar activity.

A great impact on the process of Sun’s polar field evolution has the transport of the magnetic fields over the Sun’s surface.  A special role belongs to the “Rush-to-the-Poles” flows (RTTP), which have direct relation to the polar field reversal. The RTTP phenomenon was studied in  coronal emission features in Fe XIV from the National Solar Observatory at Sacramento Peak  \citep{alt}.  Multiple Rush-to-the-Pole episodes were found in \citep{gopa}, in occurrence of high-latitude prominence eruptions. The transport of the photospheric magnetic fields (surges) was studied  in \citep{petr} and  \citep{mord}. It was found that RTTP  are the product of the decay of the following sunspots, causing a change in the sign of the polar field. In \citep{sun} magnetic flux from active regions migrating poleward and the reversal process during Cycle 24 were studied. In \citep{wang} a statistical method to analyze the poleward flux transport during Solar Cycles 21—24 was proposed. Using surface-flux transport model \citet{yeates} study the origin of a poleward surge and its contribution to the polar field in Cycle 24.

A new phenomenon which consisted in wave-like structures with periods around 2 years was described in (\citealp{vecc}; \citealp{ ulri}). In \citep{vecc} magnetic fields were studied on the base of NSO/Kitt Peak magnetic synoptic maps. The field radial component, for each heliographic latitude, has been decomposed in intrinsic mode functions through the Empirical Mode Decomposition. Poleward magnetic flux migration around the maximum and descending phase of the solar cycle was discovered which the authors connected with a manifestation of quasi-biennial oscillations (QBO). 
This result was studied in detail in \citep{ulri}, who introduced the term ``ripples’’ for such magnetic flows. In \citep{ulri} where the data of  Mt. Wilson Observatory were used, the ripples were discovered by differentiating of the time-latitude diagram. On the differentiated diagram the ripples are clearly seen, and, moreover, they exist constantly and independently of the solar activity level. A different treatment of the time-latitude diagram (deviation from the trend) also allows to obtain a new time-latitude diagram which shows the alternation of the ripples of opposite signs regardless the solar activity level. Thus, a number of similar results were obtained 
while treating the different data sets and using various methods of treatment (\citealp{vecc}; \citealp{ ulri}):  a new phenomenon was observed which manifested itself  in the emergence at low latitudes and propagation to the poles of large-scale wave-like features of magnetic field having periods of quasi-biennial oscillations (solar QBOs).
In our study \citep{vern1} we analyzed the distribution of the positive and negative magnetic fields in the photosphere using synoptic maps produced at the NSO/Kitt Peak.  Each pixel of the synoptic map was assigned a value of $+ 1$ or $-1$ in accordance with the sign of the field in this pixel. Taking into account only the sign of the magnetic field and ignoring its magnitude we observed the imbalance of positive and negative fields. The results of this study showed the cyclic alternation of magnetic field flows of the opposite polarities. 

Weak magnetic fields and their distribution on the surface of the Sun are of great interest, since they occupy a significant proportion of the solar surface. As shown by our calculations based on NSO/ Kitt Peak data  (1978--2016), $65\%$ of the solar surface are fields $|B|\leq 5$\,G and $18\%$ of the 5\,G $< |B|\leq 10$\,G; in general, $83\%$ of the total surface of the Sun is occupied by fields $|B|\leq 10$\,G.

During the solar cycle, the relative number of pixels in an interval of the field strength  does not remain constant. Two groups of fields display opposite behavior: while the number of pixels with magnetic fields from 5 to 5000 G follows the solar cycle, the weakest fields (0 to 5\,G) develop  in antiphase with the change of the solar activity. 
The same results were obtained in \citep{vern2} where instead of the pixel number we considered the time variation of fluxes for  groups of magnetic fields differing in strength. The magnetic flux of the weak magnetic fields ($|B|\leq 5$\,G) varied in antiphase with the solar cycle and with the flux of the strong fields. This is consistent with the conclusions of Jin and Wang (2014), who found that magnetic low flux structures change in antiphase with the solar cycle. 

In \citep{vern2} the time variations of the magnetic flux were compared with the variations of the nonaxisymmetrics  component of the magnetic field distribution  for magnetic fields of different strengths $B$. Unexpectedly, the variations in the nonaxisymmetric component of the magnetic field (longitudinal asymmetry) of the weak fields follow the general course of the solar cycle. Thus, the longitudinal asymmetry of the weak fields $|B|\leq 5$\,G varies in antiphase with the magnetic flux of these fields.

Time-spatial development of the weak photospheric fields was discussed in (\citealp{geta19a}; \citealp{geta19b}; \citealp{murs}). On the basis of various data sets the distribution of the weak photospheric magnetic fields was studied \citep{geta19a}. The obtained results show that  weak-field asymmetries are most pronounced in medium- and low-resolution synoptic maps being a real feature of the weak fields. The asymmetry of weak magnetic field distribution was considered separately for  the two hemispheres in \citep{geta19b} where it was found that the northern and southern hemisphere shifts are as a rule opposite to each other.

The purpose of our present work is to study in more detail the distribution of the weak magnetic fields of positive and negative polarity over the surface of the Sun. In Section 2 we discuss the used data and method of their treatment. Section 3 considers the phenomenon of ripples using time-latitude diagram both for various magnetic field values and for positive and negative magnetic fields separately. In Section 4 the main conclusions are formulated.

\section{Data and method}
	 
	  Synoptic maps of the photospheric magnetic field produced by the National Solar Observatory/Kitt Peak were used for this study. Data were obtained for 1978--2003 at ftp://nispdata.nso.edu/kpvt/synoptic/mag/) and for 2003--2016 at https://magmap.nso.edu/solis/archive.html.  Each map consisted of $180\times 360$ pixels with magnetic field values in gauss. Taking into account the sign of the magnetic field we averaged synoptic maps over longitude and constructed the time-latitude diagram which reflected the imbalance of positive and negative fields in the photosphere. In the distribution of the solar magnetic fields, especially of the weak fields ($|B|\le 5$\,G), the influence of random fluctuations of the  strength (noise) sets the limit to the accuracy of results. According to \citep{harv} this noise on the NSO/Kitt Peak synoptic maps can be as high as 2\,G per pixel in the near-pole regions. When constructing a time-latitude diagram each synoptic map was averaged over 360 longitude values. Due to this averaging, the relative contribution of random  fluctuations decreased, so that we can consider the resulting  error to be  near to the resolution of synoptic maps, i.e., of the order  of 0.1\,G.

 As a rule, strong fields are clearly seen on the time-latitude diagram in the form of Maunder butterflies, while the distribution of weak fields is poorly distinguishable. To emphasize the contribution of weak fields the transformations of synoptic maps were  made before combining separate maps into the time-latitude diagram. To study magnetic fields in the selected interval of strength from $B_{min}$ up to the $B_{max}$ two different methods of treating the data can be applied.   The first method consists in leaving unchanged only pixels with $B_{max}>B>B_{min}$. Data outside of the selected magnetic field interval are cut off by replacing the strength values  by zeros. The other method incorporates the saturation of magnetic field strength  at the certain level of B. For example, if the threshold is set at 5\,G, only pixels with fields $|B|\leq 5$\,G are left unchanged, while  larger or smaller fields are replaced by the corresponding limiting values $+5$\,G or $-5$\,G. 
The first approach will be called “cut-off ”, the second one “the saturation”.

\section{Results and discussion}

\subsection{Time-latitude diagrams of the two types}
  \begin{figure}    %%%%%%%%%%%%%%%%%% FIGURE   1
   \centerline{\includegraphics[width=0.75\textwidth,clip=]{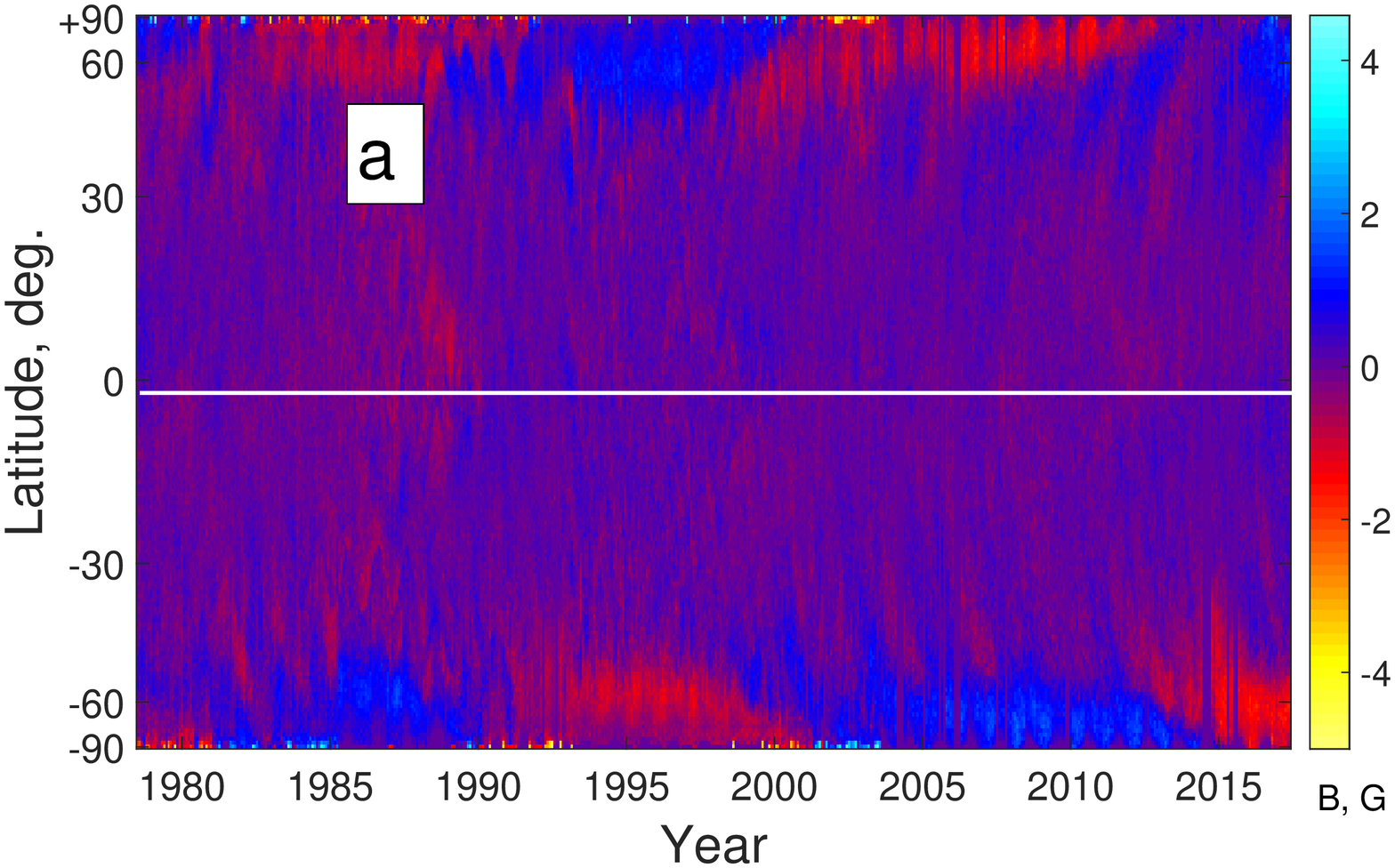}
	}
  \vspace{0.05\textwidth}
   \centerline{\includegraphics[width=0.75\textwidth,clip=]{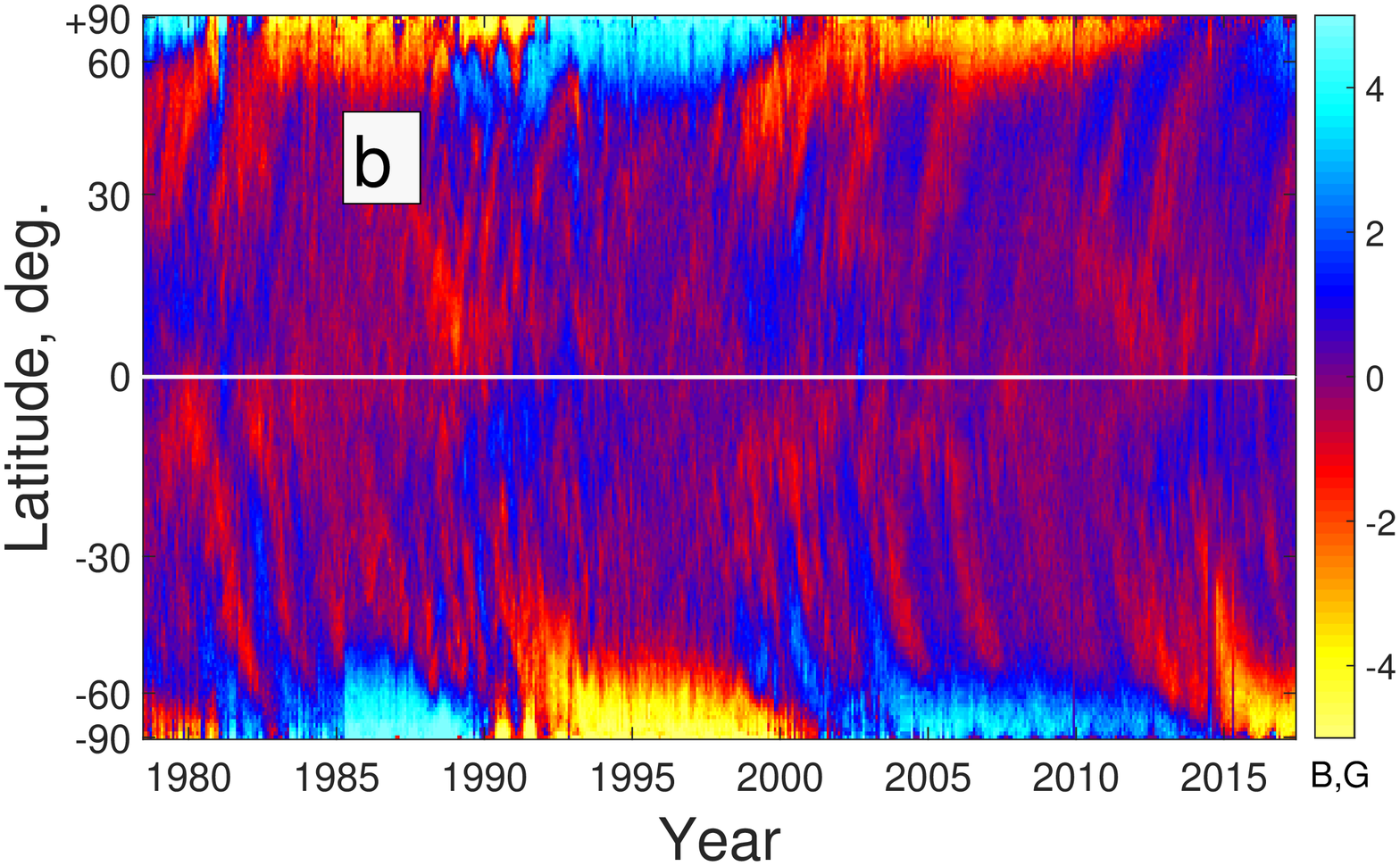}
              }
              \caption{Time-latitude diagrams of the weak magnetic field ($|B|<5$\,G) constructed in two different ways: 
							(a) pixels with $|B|>5$\,G were excluded from analysis by setting values $|B|>5$\,G in each synoptic map equal to zero. The time-latitude diagram 
              obtained by this procedure we will call "diagram with cut-off"; (b) values of  pixels with $|B|>5$\,G were replaced by limiting  
							values +5\,G or -5\,G according to the sign of $B$. Resulting diagram will be called "diagram with saturation".
                                       }
   \label{satnosat}
   \end{figure} %%%%%%%%%%%%%%%%%%%%%%%%%%%%
  
We carry out our study mostly with a time-latitude diagram built using synoptic maps with saturation at 5\,G ($|B|\leq 5$\,G). In this case pixels with magnetic field strength larger or smaller than 5\,G were replaced by the corresponding limiting values $+5$\,G or $-5$\,G. For comparison, we also plotted a diagram with cut-off at 5\,G, where all pixels with values $|B|>5$\,G were replaced by zeros. Comparison of a time-latitude diagram with cut-off (Figure~\ref{satnosat}a) and those with the saturation (Figure~\ref{satnosat}b)  shows that the cyclic change of the magnetic field polarity (ripples) can be seen in both diagrams, yet in the diagram with saturation, the picture is clearer and the details of the field distribution are more pronounced.
 
\begin{figure} [t]   %%%%%%%%%%%%%%%%%% FIGURE 2 
   \centerline{\includegraphics[width=1.0\textwidth,clip=]{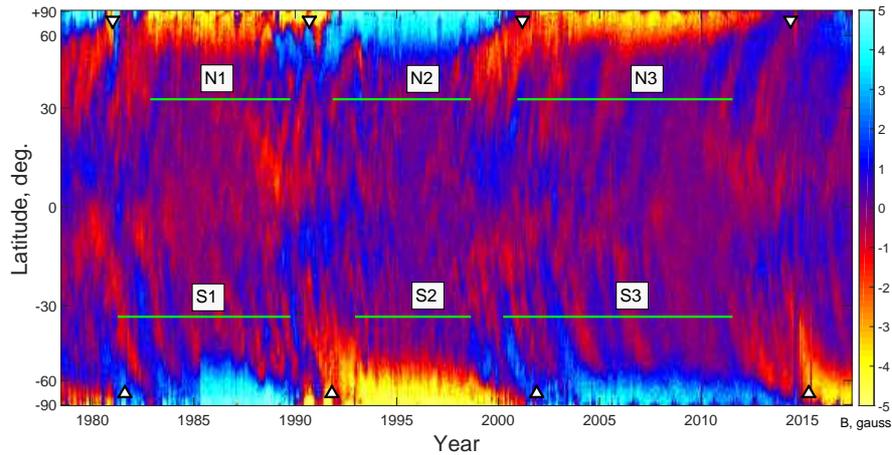}
              }
              \caption{Time-latitude diagram on the basis of synoptic maps NSO/Kitt Peak (1978--2016) with saturation at 5\,G.
							As a result, there are no Maunder butterflies and one can see alternation of bands with the dominance of a positive (blue) or negative (red) magnetic field. Cyclic variations of the magnetic field polarity can be observed associated with two types of magnetic field flows in the photosphere: Rush-to-the-Poles and ripples. Horizontal line segments point to the 6 time intervals during which the cyclic polarity alternation was observed. These intervals were denoted as N1, N2, and N3 for the northern hemisphere and as S1, S2, and S3 for the southern hemisphere. Reversals of the solar polar field marked by arrows  are taken from \citep{pishk}.            
										}
   \label{diag}
   \end{figure} %%%%%%%%%%%%%%%%%%%%%%%%%%%%

\subsection{Time-latitude diagram}
        
As described in the previous subsection, a time-latitude diagram (Figure~\ref{diag}) was obtained, in which there are no Maunder butterflies and one can see the details of the weak field distribution. The main feature of this diagram is the alternation of bands with the dominance of a positive (blue) or negative (red) magnetic field. The slope of these bands on the diagram suggests that the dominant polarity shifts in latitude over time. These bands, called ``ripples’’ in \citep{ulri}, are about one year or less wide, and can be interpreted as magnetic field flows that start near the equator and drift towards the polar regions. 

It is interesting to note that in the southern hemisphere we see a clearer pattern of alternating fields of different polarities, especially distinct for those time intervals where the polar field had a positive sign.

These ripples should be distinguished from the magnetic field flows called Rush-to-the-Poles (RTTP). Main properties of RTTP flows are given in Introduction. Magnetic fields of RTTP 
always have the  same sign  as the sign of the following sunspots and opposite to the sign of the polar field of the given hemisphere.

\begin{figure}[t]    %%%%%%%%%%%%%%%%%% FIGURE 3                                                                           
\centerline{\includegraphics[width=0.8\textwidth,clip=]{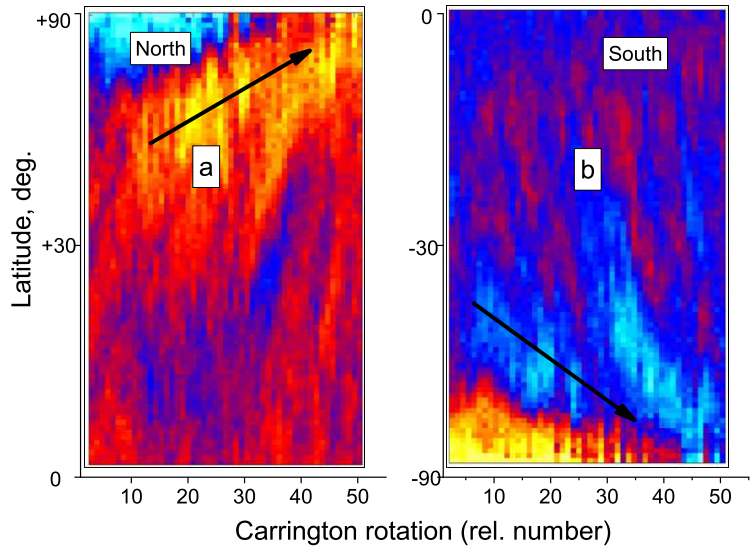}                                                     
           }                                                                                                            
           \caption{Examples of the RTTP flows in  Solar Cycle 23 (1998--2001):  (a) northern hemisphere; (b) southern hemisphere. 
					 RTTP flows begin at latitudes of $30^\circ$--$40^\circ$  and propagate to the poles. RTTP flows are about three years wide. 
					The polar field changes its sign on arrival of RTTP to the polar regions. The arrows indicate 
					the direction of latitudinal movement of flows in time.          
                       }                                                                                                
\label{rttp}                                                                                                            
\end{figure} %%%%%%%%%%%%%%%%%%%%%%%%%%%%     
                                                                     
\begin{figure}    %%%%%%%%%%%%%%%%%% FIGURE 4                                                                           
\centerline{\includegraphics[width=0.8\textwidth,clip=]{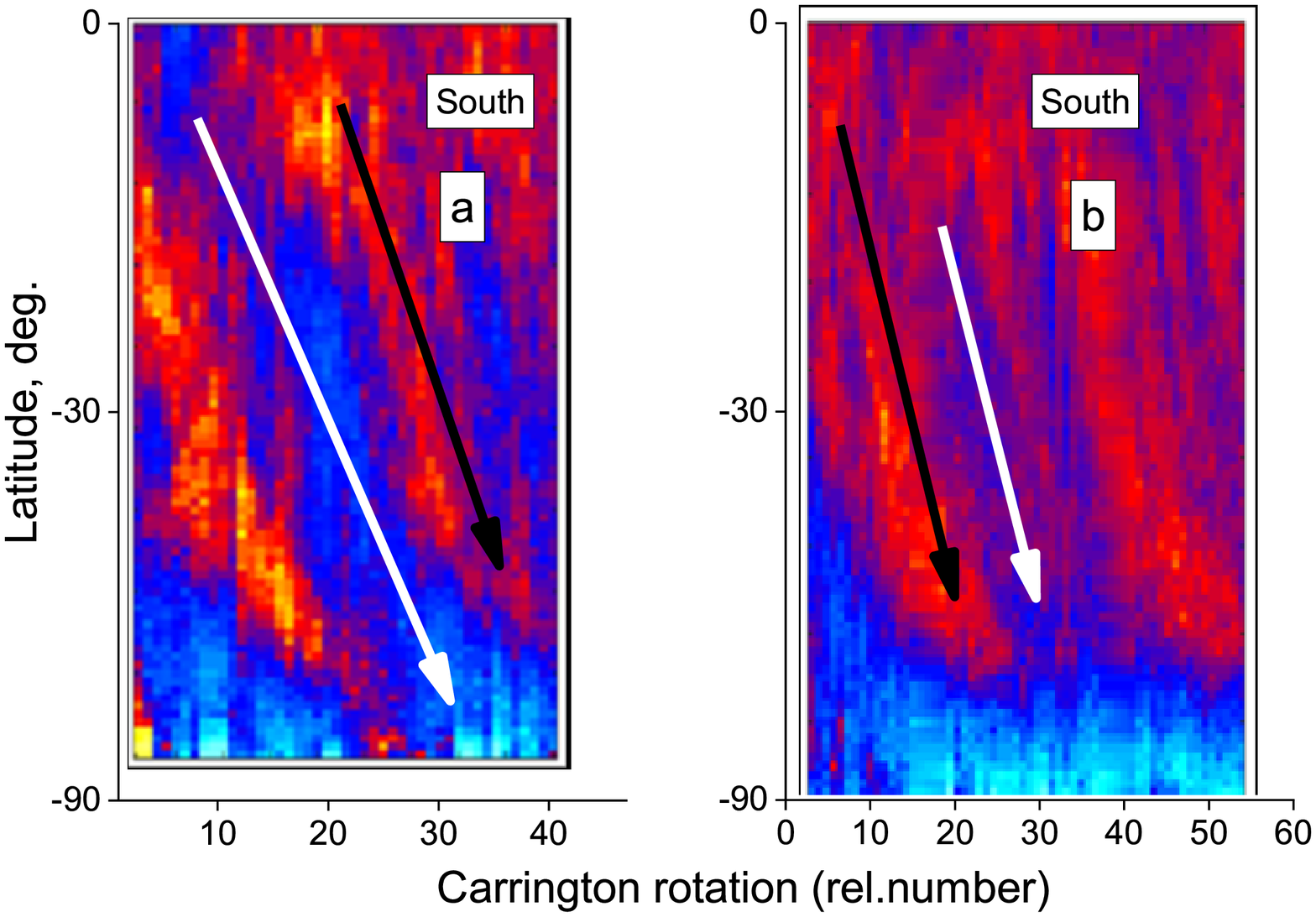}                                                  
           }                                                                                                            
           \caption{Examples of ripples in southern hemisphere (a) 1981--1984; (b) 2003--2007. Ripples in contrast to RTTP are periodic 
					structures consisting of a set of fluxes with the opposite polarities. The width of separate flux in the set 
					is from 0.5 yr (a) to 1 yr (b). Ripples appear near the equator and reach latitudes $~50^\circ$. In this interval, 
					the ripples field sign  changes four times, while the field sign in RTTP remains unchanged.	The arrows indicate 
					the direction of latitudinal movement of flows in time.          		                                            
                                    }                                                                                   
\label{ripp}                                                                                                            
\end{figure} %%%%%%%%%%%%%%%%%%%%%%%%%%%%
On the time-latitude diagram (Figure~\ref{diag}), RTTPs are clearly visible near the time of solar maxima as rather large bands about 2--3 years wide, 
which begin at latitudes $\sim 30^\circ$--$40^\circ$. The times when these streams reach the poles coincide with the reversals of the polar field. 

Unlike RTTPs, which have the form of separate magnetic field flows appearing once per solar cycle at its maximum, ripples are a set of narrow field flows with alternating polarity. Each series of ripples spans a period of about 10 years, including  decrease, minimum, and rising phases of the solar cycle. We observe such flows looking like a wave packets in the time interval between two RTTPs, when the sign of the polar field is constant. In contrast to this result, \citet{vecc} found similar structures only in years of high solar activity. Another conclusion was made in \citep{ulri} where similar flows were registered during all phases of solar cycle.
                                                                  
In the data array under consideration represented by diagram of Figure~\ref{diag} we selected 6 time intervals between neighboring RTTPs: 3 in the northern hemisphere (N1, N2, N3) and 3 in the southern hemisphere (S1, S2, S3). In these intervals, the alternation of the flow polarity (ripples) can be clearly seen.

To illustrate the difference between RTTP and ripples the cuts from the time-latitude diagram are presented in Figure~\ref{rttp} (RTTP) and Figure~\ref{ripp} (ripples). RTTP flows are shown for the  Solar Cycle 23 for the northern hemisphere (Figure~\ref{rttp}a) and for the southern hemisphere (Figure~\ref{rttp}b). As an example of ripples, we chose two sections of the time-latitude diagram in the southern hemisphere, both selected sections (Figures~\ref{ripp}a,b) falling on the period of time when the sign of the polar field was positive. The arrows in Figures~\ref{rttp},~\ref{ripp} indicate the direction of latitudinal movement of flows in time.

Magnetic field flows forming ripples (Figure~\ref{ripp}) appear, as a rule, near the equator and reach latitudes $\sim 50^\circ$. In contrast, RTTP flows (Figure~\ref{rttp}) begin at latitudes of $30^\circ$--$40^\circ$ and propagate to the poles.

The width of the flow (the duration of constant field sign at a fixed latitude) is noticeably different for ripples and RTTP. The width of the flows with positive or negative polarities forming the ripples ranges from 0.5 years (Figure~\ref{ripp}a) to 1\,year (Figure~\ref{rttp}b). Thus, the total period including the flows of both polarities is approximately 1--2 years. RTTP flows are significantly wider (Figure~\ref{rttp}) -- their width is about three years. We estimated the average life time of the RTTP from 8 streams at latitudes $+ 50^\circ$ of the northern hemisphere and $-50^\circ$ of the southern hemisphere. This time  proved to be $3.2 \pm 0.3$ years).

 The  selected sections of the time-latitude diagram  in Figures~\ref{rttp},~\ref{ripp} have a time length of  $\sim 50$ Carrington rotations. In this interval, the ripples field sign (Figure~\ref{ripp}) changes four times, while the field sign in RTTP remains unchanged (Figure~\ref{rttp}).

  \begin{figure}    %%%%%%%%%%%%%%%%%% FIGURE   5
   \centerline{\includegraphics[width=0.95\textwidth,clip=]{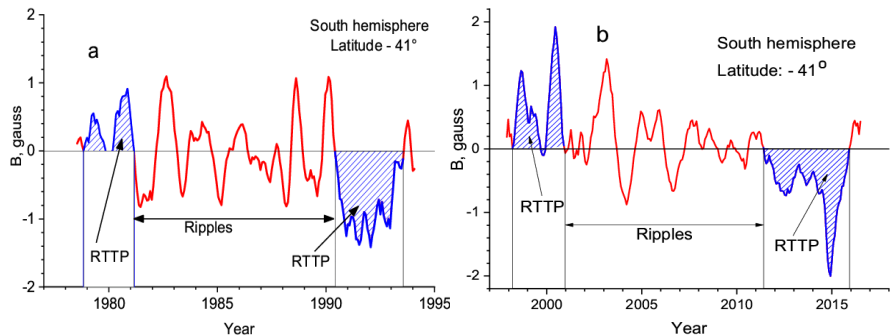}
              }
              \caption{ Change of the magnetic field along the latitude $41^\circ$ of the southern hemisphere (see Figure~\ref{diag}). 
              RTTP are marked with shading. The section with cyclic variations of the magnetic field (S1) is located between two RTTPs. 
              The data are smoothed by running average over 5 points.            
													}
   \label{row}
   \end{figure} %%%%%%%%%%%%%%%%%%%%%%%%%%%%

Both ripples and RTTPs can be observed in the same plot displaying time change of the magnetic field at some selected latitude. Figure~\ref{row}  shows the magnetic field variations at the latitude of $41^\circ$ of the southern hemisphere for two intervals: 1978--1994 (Figure~\ref{row}a) and   1998--2015  (Figure~\ref{row}b). It can be seen that the periodical change of the magnetic field polarity (ripples) persists between two successive RTTPs.  This variation was  observed for 9 years in Figure~\ref{row}a and for 10 years in Figure~\ref{row}b. During these periods maxima and minima of the magnetic field curve in Figure~\ref{row} corresponded to positive and negative magnetic fields of the ripples. 

The RTTP flows are marked in Figure~\ref{row} by a shading. The first RTTP in each of the two pairs  (Figures~\ref{row}a,b) has the positive magnetic field polarity which is opposite to the sign of the polar magnetic field and coincides with the polarity of the following sunspots. Accordingly, the pair of the second RTTP in Figures 4a,b displays negative polarity just as the field of the following sunspots.

\begin{figure}    %%%%%%%%%%%%%%%%%% FIGURE  6 
   \centerline{\includegraphics[width=0.95\textwidth,clip=]{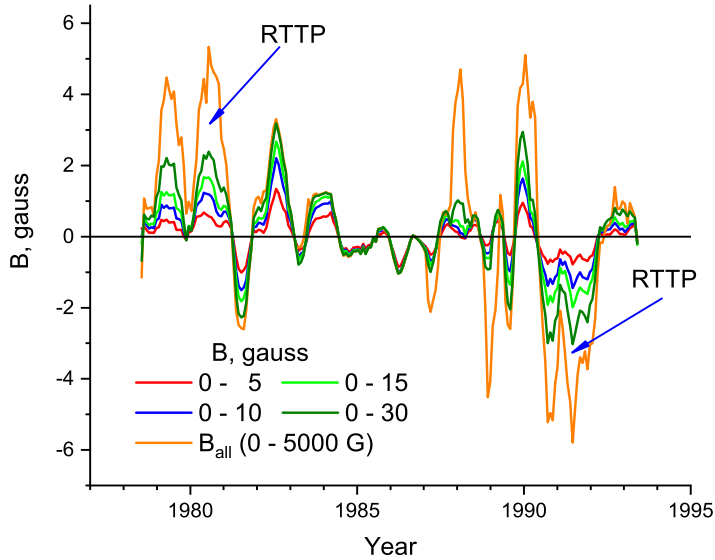}
              }
              \caption{Influence of the saturation limit on the change of magnetic field along the latitude $33^\circ$ of the southern hemisphere. The saturation thresholds are set at  $B=5, 10, 15, 30$ and $B=5000$\,G; the last one means inclusion of all magnetic fields. One can see that irrespective of the saturation thresholds magnetic fields change synchronously.
																									}
   \label{b10}
   \end{figure} %%%%%%%%%%%%%%%%%%%%%%%%%%%%

To compare the variations of the magnetic fields with different strength we plotted time-latitude diagrams using various saturation thresholds ($B_{lim}= 5, 10, 15, 30, 5000$\,G). For these diagrams change of magnetic field along the latitude $33^\circ$ of the southern hemisphere is shown in Figure~\ref{b10}. One can see that all curves develop synchronously yet their behavior during the  phases of low and high solar activity is different. From one RTTP to another cyclic change of the magnetic field polarity can be seen (ripples).  During 4 years of the low solar activity most of the variation is connected with the weak fields ($B\leq 5$\,G). The increase of the threshold value does not produce a significant rise of the magnetic field variation. On the contrary, 
for high solar activity periods (where RTTP are present) the increase of the selected threshold leads to a significant rise of the variation. The highest values of $B$ are observed when the threshold is set at 5000\,G which is equivalent to the inclusion of all magnetic fields. Strong fields  change their sign at the same time moments as the weakest fields. It means that there is a common pattern in the alternation of the field signs for all strengths  of magnetic fields. Due to the synchronous behavior of magnetic fields of different strengths these fields can be combined into the same time-latitude diagram.

\subsection{Polarity variations for different strengths of magnetic field}

When constructing the time-latitude diagram (Figure~\ref{diag}) for each synoptic map, fields with the modulus greater than 5\,G were replaced by the limit values of $+5$\,G and $-5$\,G. This procedure was applied to each of the synoptic maps which were then longitude averaged and included in the time-latitude diagram. In this way the influence of the strong magnetic fields was suppressed revealing the periodic variations of the low strength fields. The strong fields were also accounted for in construction of the time-latitude diagram yet with the saturation limit of  $|B| = 5$\,G. 

The relative contribution of the different field groups to the distribution of magnetic fields varies according to the field strength. In this connection, the question arises which magnetic fields play the main role in emergence of a periodic structure in the form of the alternating bands of different polarity.

To answer this question we considered the following groups of fields: 0--5\,G, 5--15\,G, 15--50\,G, and $|B|>50$\,G. For the construction of the time-latitude diagrams the transformed synoptic maps were used where only field values in the selected intensity ranges were left unchanged. All pixels not included in these groups of fields were filled with zeros.

\begin{figure}    %%%%%%%%%%%%%%%%%% FIGURE 7 
  \centerline{\includegraphics[width=0.95\textwidth,clip=]{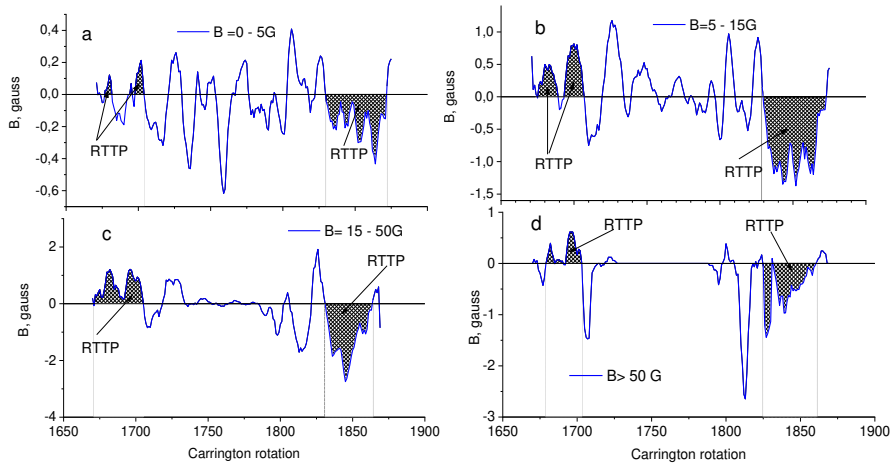}
              }
              \caption{Contribution of fields of different strength to cyclic variations. 
							Change of the magnetic field along the latitude $41^\circ$ of the southern hemisphere 
							is shown for each of the field groups. RTTPs are marked with shading.
							The data are smoothed by running average over 5 points.
                      }
   \label{abcd}
   \end{figure}

For these four field groups the changes of magnetic fields at the latitude $41^\circ$ of the southern hemisphere 
are shown in Figure~\ref{abcd}. Periodic reversal of the magnetic field sign is clearly seen for the 
fields of 0--5\,G (Figure~\ref{abcd}a), and 5--15\,G (Figure~\ref{abcd}b),  in the interval from one Rush-to-the-Pole to another. For the intermediate group of 15--50\,G (Figure~\ref{abcd}c),   the alternation of polarities occurs only at a sufficiently high level of solar activity, and the variations disappear near the minimum. Periodic polarity changes are almost completely absent for fields 
with $|B| > 50$\,G (Figure~\ref{abcd}d). Thus, the alternation of the dominant polarity of flows is a characteristic property of weak fields.

As can be seen in Figure~\ref{abcd}, Rush-to-the-Poles flows (shaded in the figure) are present in all field groups. Thus, in the formation of Rush-to-the-Poles flows, not only weak fields of less than 15\,G are involved, but also fields from 15 to 50\,G and more.

On the other hand, the synchronous behavior of different field groups allows their inclusion in the time-latitude diagram while studying the features of ripples. In order to avoid excessive influence of the most strong fields the saturation limit should be  chosen and set at sufficiently low level.

 \begin{figure}    %%%%%%%%%%%%%%%%%% FIGURE    8
   \centerline{\includegraphics[width=1.0\textwidth,clip=]{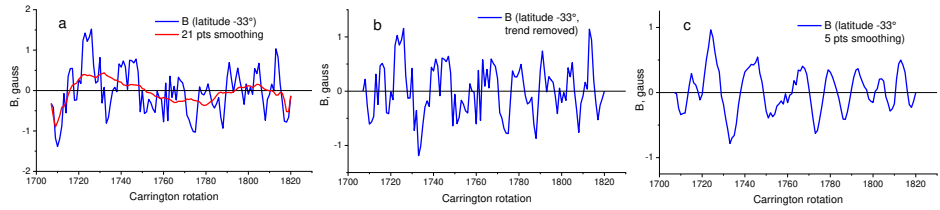}
              }
              \caption{(a) Change of the magnetic field ($|B| \le 5$\,G) along the latitude  $-33^\circ$ 
              in the interval S1 (blue curve). The trend was determined as a result of running smoothing 
              over 21 points (red curve); (b) magnetic field after detrending; 
              (c) magnetic field (b) smoothed over 5 points.
                                       }
   \label{prim}
   \end{figure} %%%%%%%%%%%%%%%%%%%%%%%%%%%%

\subsection{Period and amplitude of variations}

In the time-latitude diagram (Figure~\ref{diag}), which served as an experimental basis for our analysis of magnetic field variations, flows with different polarity appear very clearly. However, upon closer examination, it turns out that this phenomenon is quite complex, and over several cycles the variation parameters experience significant changes. To begin with, the lifetime of these variations (the section of the time-latitude diagram in which the variation is constantly present) changes during 1978--2016 from 6 to 11 years in the northern and southern hemispheres. Approximate estimates of the period and amplitude of the variations also showed that these parameters differ significantly between sections N1, N2, N3, S1, S2, S3. Therefore, to obtain quantitative estimates of the period and amplitude parameters, it was necessary to subject the time-latitude diagram to a preliminarily processing that would eliminate both the shortest variations and long period ones from the primary data, since it was obvious that ripples are associated with variations that have a period  from one year to several years
      \begin{figure}    %%%%%%%%%%%%%%%%%% FIGURE    9
   \centerline{\includegraphics[width=0.95\textwidth,clip=]{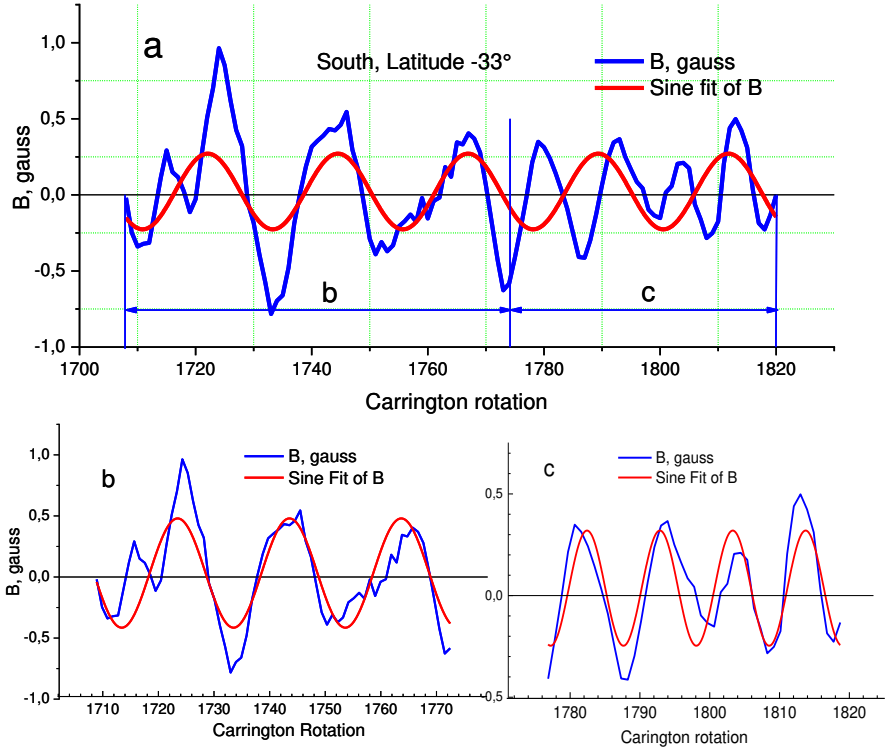}
              }
              \caption{Approximation of magnetic field variations by a sinusoidal function: 
              (a) for the entire interval S1; (b) and (c) -- the approximation interval is 
								divided into two parts. 
              There is a noticeable difference in the periods of variations 
							for parts (b) and (c).
                                 }
   \label{tpar}
   \end{figure} %%%%%%%%%%%%%%%%%%%%%%%%%%%%

To isolate these variations, the following data processing technique was adopted. As an example, we consider the change of the magnetic field along the latitude $-33^\circ$ in the interval S1 (Figure~\ref{prim}a, blue curve). The trend of the raw data (Figure~\ref{prim}a, red curve) was determined as a result of a running smoothing over 21 points (a point corresponds to one Carrington rotation). After subtracting the trend (Figure~\ref{prim}b), the data was smoothed over 5 points (rotations). As a result of such processing, both the slowest and fastest variations of the field were excluded from the data series, and thus the interval of periods of interest was emphasized and we could observe the variation in a ``pure'' form (Figure~\ref{prim}c). The time dependences of the magnetic field strength at the latitude of $-33^\circ$ were subjected to such processing for 6 intervals marked on the latitude-time diagram (Figure~\ref{diag}).

After that, it became possible to estimate the amplitude and period by fitting the ``cleaned data'' 
with a sinusoidal function. The magnetic field variations for the intervals N1, N2, N3, S1, S2, S3 were 
approximated by a function of the following form:
\begin{equation}
\label{sinap}
			y=y_0+A \sin \frac{2 \pi (t-tc)}{T}, 
\end{equation}
where $A$ is the amplitude, $T$ is the period of variation and $tc$ is the shift of the phase.
(Figure~\ref{tpar} shows the approximation for the interval S1). It turned out that the variation period noticeably changes not only from one interval to another, but also within one interval, i.e., over a time of $\sim{10}$ years (in our example, during the interval S1), which leads to uncertainty in the estimation of the period. Therefore, the intervals were divided into two parts (with a split point near the minimum of the solar cycle) and the approximation was performed independently for each of the parts (see Figure~\ref{tpar}b,c).

\begin{figure}    %%%%%%%%%%%%%%%%%% FIGURE     10
   \centerline{\includegraphics[width=0.95\textwidth,clip=]{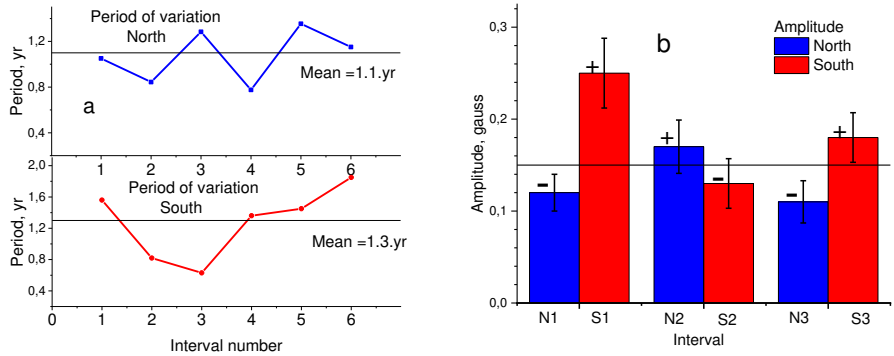}
              }
              \caption{Periods and amplitudes of the magnetic field variations (ripples) along the $33^\circ$  latitudes in the northern and southern hemispheres. 
							(a) Periods of variations for both hemispheres. The  six intervals N1, N2, N3, S1, S2, S3 were divided in two equal parts; 
							the  periods were evaluated independently for each of the  12 sections. (b) Amplitudes of the variations for 6 time intervals (N1--S3). 
							At the top of histogram the sign of the polar field is shown. The amplitudes were higher for the hemisphere with positive polar field.
                                       }
   \label{hist}
   \end{figure}

Figure~\ref{tpar}b,c shows that approximating the two parts separately gives a better accuracy. The data of all 6 intervals were processed similarly and the period and amplitude of the variations were determined. This analysis showed that the variation period changes during the N1--S3 intervals, and in the northern hemisphere, in each interval before the minimum of solar activity, the variation period was longer than that after the minimum (Figure~\ref{hist}a). However in the southern hemisphere, such effect appeared only in one of the three regions. 
In average, the period of variation was 1.1 years for the northern hemisphere and 1.3 years for the southern hemisphere.
  %%%%%%%%%%%%%%%%%%%%%%%%%%%%

It should be emphasized that these estimates of the period of variation are obtained for particular latitudes ($+33^\circ$ and $-33^\circ$). For other latitudes, we received slightly different periods. 
In the work \citep{vern3}, averaged period values for the range of latitudes from the equator to $50^\circ$ were obtained by two methods: a) using the method of empirical orthogonal function (EOF) analysis and b) by summing the time profiles of the field taking into account their latitude shift with time.The values  of the periods for two methods turned out to be: 1.8 years and 1.6 years. 

These  period values are close to the estimations of other authors. In \citep{vecc} these variations are considered as one of manifestations of the  quasi-biennial oscillations (QBO). Period of ripples  from 0.8 to 2 years was found  in \citep{ulri}.

A certain regularity is seen in the change of the amplitude of the variations of the two hemispheres (Figure~\ref{hist}b): the amplitude is higher for the hemisphere in which the polar field is positive (the sign of the field is indicated in the upper part of the histogram). This effect supports the connection of polarity variations with the 22-year magnetic cycle of the Sun.

\begin{figure}    %%%%%%%%%%%%%%%%%% FIGURE     11
   \centerline{\includegraphics[height=0.7\textwidth,width=0.95\textwidth,clip=]{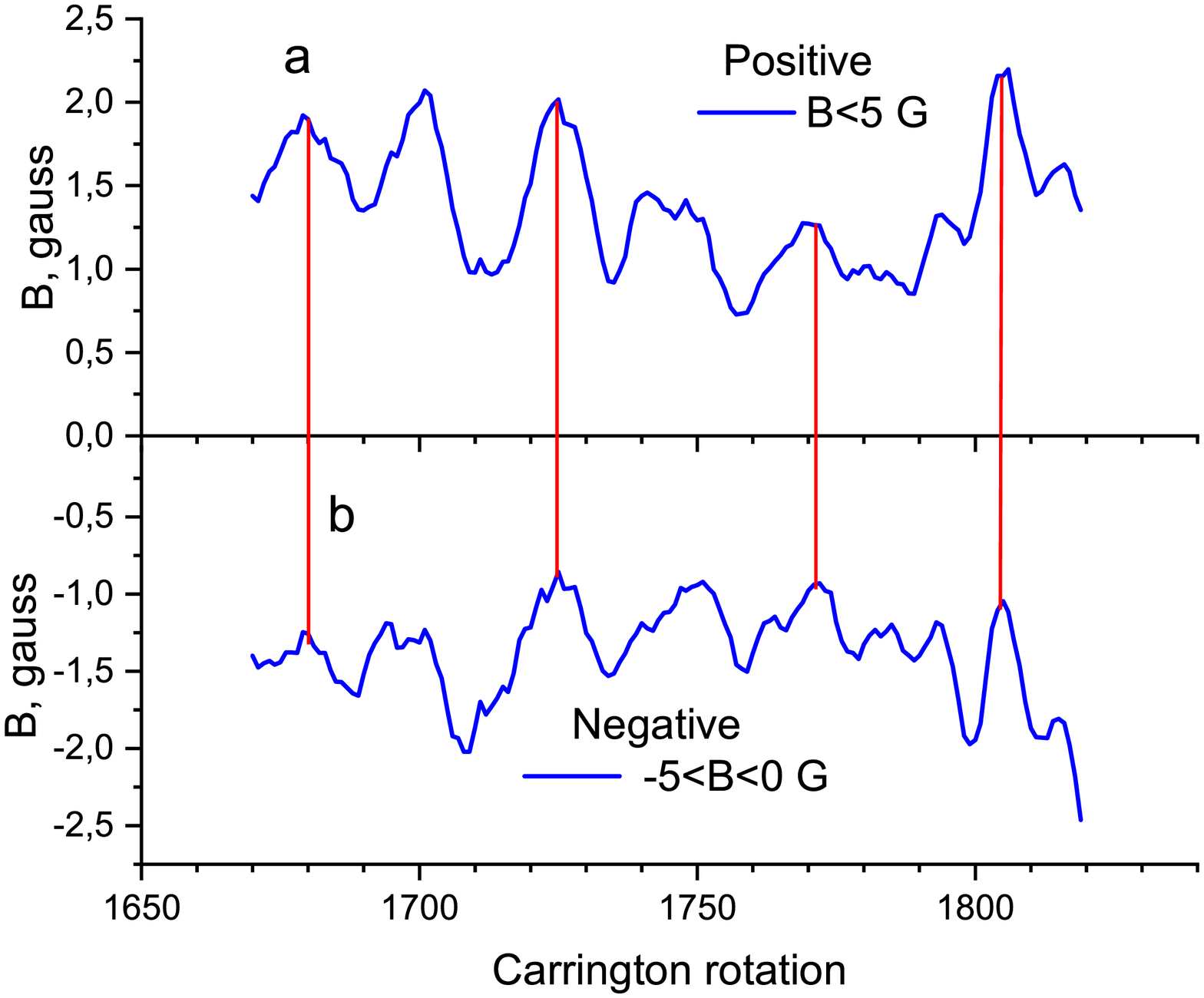}
              }
              \caption{Development of positive and negative magnetic fields considered separately. Time-latitude diagrams were constructed 
							for $0< B \leq 5$\,G and $-5\leq B <0$\,G. Magnetic field for the interval S1 at the latitude $-33^\circ$:  (a) only positive field; (b)  negative field. 
							Red lines show the relative positions of the extrema of the two curves.
                                       }
   \label{pone}
   \end{figure} %%%%%%%%%%%%%%%%%%%%%%%%%%%%

    \begin{figure}    %%%%%%%%%%%%%%%%%% FIGURE     12
   \centerline{\includegraphics[height=0.7\textwidth,width=0.95\textwidth,clip=]{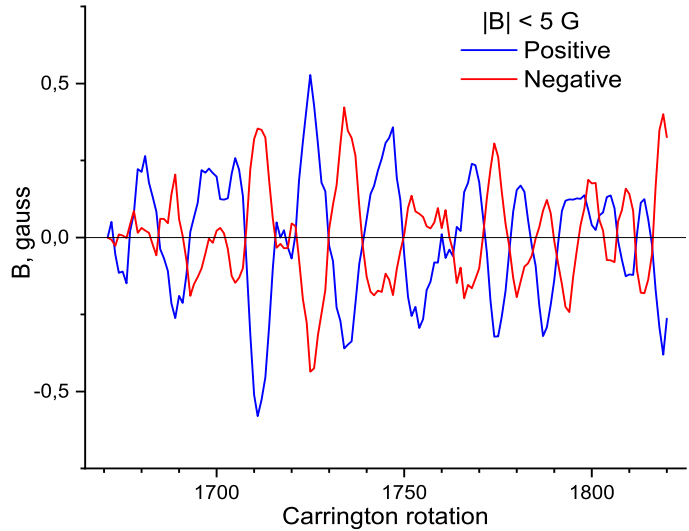}
              }
              \caption{Comparison of variations of positive (blue curve) and modulus of negative (red curve) magnetic fields. 
							The same interval as in Figure~\ref{pone} is displayed  after removing the trend (running average for 21 Carrington 
							rotations). The fields develop strictly in anti-phase.
                                       }
   \label{pabsn}
   \end{figure} %%%%%%%%%%%%%%%%%%%%%%%%%%%%
\subsection{Fields of positive and negative polarity}

The contribution of fields of positive and negative polarity to the formation of a cyclic structure of flows with alternating signs of the field is considered.

There are two possibilities for the magnetic field variations, which lead to a variation of the dominant polarity. The first variant: positive fields and the modulus of negative fields develop in phase with each other, but there is a cyclical change of the ratio between their values. Another variant is that high values of positive fields correspond to low (in the modulus) negative fields and vice versa. The result will be the same: alternate dominance of one of the polarities.

Plotting the time-latitude diagrams separately for positive and negative fields, one can check which of the variants takes place. We used the same value of 5\,G as the saturation limit for synoptic maps.

Figure~\ref{pone} shows the time variation of positive and negative magnetic fields at the latitude $33^\circ$ of the southern hemisphere. Within the same flow, fields of different signs are closely related to each other and develop in anti-phase: the maxima of the positive field (upper curve) are close in time to the minima of the absolute value of the negative field (lower curve). (These points are connected by red lines). To quantify this effect, a running smoothing of positive magnetic fields over 21 points was performed. The same smoothing was performed for the modulus of negative fields.

If we subtract the 21-point smoothed values (trend) from the positive values and from the absolute negative values and then consider the correlation between the positive and the negative values, then it will be high ($R= - 0.83$) for the interval S1 of 150 rotations and even higher ($R= -0.89$) for a smaller interval of 100 rotations.

Thus, an increase in the positive field within one flow  is accompanied by a decrease in the modulus of the negative field in the same flow  and vice versa. This leads to alternate dominance of magnetic fluxes with different polarities.

\section{Conclusions}

Wave-like structures with periods around 2 years (the ripples) were found in  (\citealp{vecc}, \citealp{ulri}). We continued the study of this phenomenon using the time-latitude diagrams constructed on the base of NSO/Kitt Peak data. To emphasize the contribution of weak fields the following transformation of synoptic maps was made: for each synoptic map only magnetic fields with modulus less than 5\,G ($|B|\leq 5$\,G) were left unchanged while  larger or smaller fields were replaced by the corresponding limiting values $+5$\,G or $-5$\,G. A time-latitude diagram was obtained, in which there are no Maunder butterflies and one can see the details of the weak field distribution.
In the time-latitude diagram, one can see magnetic field fluxes of two types: Rush-to-the-Poles (RTTP) and the ripples. The two phenomena have very different characteristics.
Magnetic fields of RTTP always have the  same sign  as the sign of the following sunspots and opposite to the sign of the polar field of the given hemisphere. The times when RTTP reach the poles coincide with the reversals of the polar field. Unlike RTTP, which appear as separate magnetic field flows one per maximum of the solar cycle, ripples are a set of narrow field flows with alternating polarity. We observe such flows in the time interval between two RTTPs, when the sign of the polar field is constant, i.e., at the decrease, minimum and rise phases of the solar cycle. 
RTTP flows begin at latitudes of $30^\circ$--$40^\circ$ and propagate to the poles. In contrast, magnetic field flows forming ripples appear, as a rule, near the equator and reach latitudes $\sim 50^\circ$. 
The width of RTTP flows  is about three years.  The width of the flows with positive or negative polarities of the ripples ranges from 0.5 years  to 1 year.
The main contribution to the ripples is made by weak fields $|B| \le 15$\,G. Fields from 15 G to 50 G and more are involved in formation of RTTP flows.
Two types of flows -- RTTP and ripples -- together form a structure which appears regularly in the magnetic field of the photosphere and has close connection to the solar magnetic cycle.

%% Acknowledgements
%
\begin{acks}
The NSO/Kitt Peak data used here are produced cooperatively by NSF/NOAO, NASA/GSFC, and NOAA/SEL (ftp://nispdata.nso.edu/kpvt/synoptic/mag/). Data acquired by SOLIS instruments 
were operated by NISP/NSO/AURA/NSF (https://magmap.nso.edu/solis/archive.html). 
\end{acks}

%
% \begin{ethics}
% \begin{conflict}
%
% \end{conflict}
% \end{ethics}

%%% %%%%%%%%%%%%%%%%%%%%%%%%%%%%%%%%%%%%%%%%%%%%%%%%%%%%%%%%%%%
%% Bibliography
%
% Using BibTeX
%
% \bibliographystyle{spr-mp-sola}
% \bibliography{<bib file>}  
%
% Without BibTeX 
% \begin{thebibliography}{}
% \bibitem[\protect\citeauthoryear{Author}{Year}]{key}
%   <bibliographical entry>
%
% \bibitem[\protect\citeauthoryear{}{}]{}
%   
%  
% \end{thebibliography}

\end{article} 
\end{document}